\documentstyle[preprint,aps]{revtex}
\begin{document}
\draft
\title{Raman scattering spectra of elementary electronic excitations
\protect\\ in coupled double quantum well structures}
\author{P.\ I.\ Tamborenea and S.\ Das Sarma}
\address{Department of Physics \\
         University of Maryland \\
         College Park, Maryland  20742-4111}
\date{\today}
\maketitle
\begin{abstract}
Using the time-dependent-local-density-approximation (TDLDA)
within a self-consistent
linear response theory, we calculate the elementary excitation energies
and the associated inelastic light scattering spectra of a strongly coupled
two-component plasma in a double quantum well system with electron occupation
of symmetric and antisymmetric subbands.
We find, consistent with the results of a recent experimental Raman scattering
study, that the intersubband spin density excitations tend to merge with the
single particle excitations (i.e. the excitonic shift decreases
monotonically) as the Fermi energy increases beyond the
symmetric-antisymmetric energy gap $\bigtriangleup_{SAS}$.
However, our TDLDA calculation does not show the abrupt suppresion
of the excitonic shift seen experimentally at a finite value
of the subband occupancy parameter
$\eta \equiv \bigtriangleup_{\text{SAS}} / E_{\text{F}}$.
\end{abstract}

\pacs{73.20.Mf, 71.45.Gm, 78.20.Ls, 78.66.Fd}

\narrowtext

In an interesting recent experimental Raman scattering study, Decca
{\it et al.} reported \cite{decca} the observation of the suppression
of the collective
intersubband spin density excitation (SDE) in a coupled double quantum well
(DQW) structure in the high electron density ($N_s$) limit when the two
low-lying DQW subbands, the so-called symmetric (S) and antisymmetric
(AS) levels with a single-particle energy gap
$\bigtriangleup_{SAS} < E_{\text{F}}$
(where $E_{\text{F}}$ is the two dimensional Fermi energy)
separating them, are both densely occupied.
In this Rapid Communication we provide a detailed quantitative calculation
of the elementary excitation spectra of the coupled DQW system discussing
in particular the recent inelastic light scattering experimental
observations.
We find excellent agreement between our theoretical results and experimental
data for both the charge density excitation (CDE) and the SDE spectra except
for the samples with highest $N_s$ where the subband filling parameter
$\eta\equiv\bigtriangleup_{\text{SAS}}/E_{\text{F}}\stackrel{<}{\sim}0.25$.
Our main interest in this paper, following Ref.\ \onlinecite{decca}, is to
investigate the strong-coupling two-component situation $\eta < 1$
when the S and AS subbands are both occupied by electrons.
Around a small critical value of $\eta$ ($\approx 0.1$), the experimental
SDE abruptly merges with the single particle excitation (SPE) spectra,
whereas our calculated SDE merges with the SPE monotonically as a continuous
function of $\eta$ as $\eta$ approaches zero without showing any sudden
collapse around $\eta \approx 0.1$.
Aside from this important qualitative difference associated with the
abrupt collapse of the experimental vertex correction (the energy
difference between the SPE and the SDE arises from the excitonic
vertex correction) around $\eta \approx 0.1$, our theoretical
results agree very well (within 0.5 meV) with the experimental measurements.

A typical DQW structure used in our calculation is shown in Fig.\ 1, where
the electron gas is taken to be confined in the translationally invariant
x-y plane and the growth direction is the z-axis.
Following the experimental work of Decca {\it et al.} \cite{decca}
we have used in our
calculations several different DQW structures with varying
$N_{\text{s}}$ and well parameters so as to have many different
values of $\bigtriangleup_{\text{SAS}}$ and the occupancy parameter
$\eta$ ($\equiv \bigtriangleup_{\text{SAS}}/E_{\text{F}}$).
Our calculation involves three steps.
First, we carry out a self-consistent local density approximation
(LDA) calculation \cite{ando,ste-das,mar-das} of the DQW subband energy
levels $E_i$
and wavefunctions $\phi_i(z)$ (a typical example being shown in Fig.\ 1)
using the best available exchange-correlation potential \cite{cep}.
Note that by definition $\bigtriangleup_{\text{SAS}}=E_2 - E_1$ is the
energy difference
between the lowest two subbands of the DQW.
Then we calculate the irreducible ($\Pi$) and the reducible
($\tilde{\Pi}$) polarizability functions of the electron layer
confined in the DQW structure using the self-consistent LDA subband energies
and wavefunctions in a linear response theory
\cite{ando,mar-das,das2}.
The irreducible and the reducible response functions are formally
connected by the matrix relation $\tilde{\Pi}=\Pi \varepsilon^{-1}$
where $\varepsilon=1 - V \Pi$ is the tensor representing the subband
dielectric function with V as the direct Coulomb interaction.
The irreducible response function $\Pi$ is connected to the bare LDA
polarizability function $\Pi_0$ (which is just the two-dimensional
Lindhard function using LDA energies and wavefunctions) through the vertex
correction $\Pi = \Pi_0 (1+U_{\text{xc}} \Pi)^{-1}$ where $U_{\text{xc}}$
represents the LDA exchange-correlation induced vertex correction.
Thus $\Pi_0$, $\Pi$, and $\tilde{\Pi}$ are respectively the ``bare''
polarizability bubble (including, however, renormalized LDA quasiparticle
energies), the vertex corrected polarizability, and the screened
polarizability.
Once $\Pi_0$, $\Pi$, and $\tilde{\Pi}$ are calculated their respective
poles (or, peaks) immediately give us the SPE, the SDE, and the CDE energies.
The third step of our calculation involves directly obtaining the Raman
scattering spectra for the SDE and the CDE which are given respectively
by the spectral strengths $Im \; \Pi$ and $Im \; \tilde{\Pi}$.
(For the purpose of comparison we also calculate the SPE spectral strength
given by $Im \; \Pi_0$, which according to the simple linear response
theory, should not be accessible to Raman scattering experiments \cite{decca}.)
Some typical calculated Raman spectra are shown in Fig.\ 2 of this paper.
The self-consistent linear response integral equations connecting
the dynamical polarizability functions $\Pi_0(z,z')$, $\Pi(z,z')$,
$\tilde{\Pi}(z,z')$, are solved in the subband representation.
The procedure is standard and has been described in the literature
\cite{mar-das,das2}.
We adapt this technique for the experimental DQW structures of
Ref.\ \onlinecite{decca}
using the self-consistent LDA wavefunctions and energies.
In the rest of this paper we discuss our numerical results comparing
critically with the experimental data \cite{decca}.
All our calculations assume the effective mass approximation for the
$\text{GaAs}-\text{Al}_x \text{Ga}_{1-x} \text{As}-\text{GaAs}$
DQW structures and we use GaAs conduction band parameters
\cite{ste-das} in our LDA calculations.
The DQW well parameters and electron gas densities are taken from
Ref.\ \onlinecite{decca}.

In Fig.\ 2 we show calculated results for the spectral weights of the
SDE and SPE modes for several values of the wavevector transfer $q$ for
a sample with $N_{\text{S}}=6.35\times10^{11}cm^{-2}$ and $\eta=0.12$.
When confronted with Fig.\ 2 of Ref.\ \onlinecite{decca}, this figure
shows that the collapsed SDE and SPE experimental peaks show a wave-vector
dependent lineshape that agrees better with the theoretical lineshape
of the SPE than with that of the SDE.
Aside from the results shown in Fig.\ 2, we only discuss results
corresponding to the experimental backscattering
geometry with very small wavevector transfer, $q=0.1 \times 10^4 cm^{-1}$
as used in the experimental set-up \cite{decca}, in the plane of the
electron layer.
(All the elementary excitation energies presented in this paper correspond
essentially to the long wavelength, i.e. $q \approx 0$, limit for the
mode wavevector in the x-y plane because this is the situation reported
in the work of Ref.\ \onlinecite{decca}).
The results of this paper correspond to {\em intersubband} elementary
excitations (associated with the quantized z-motion of the electron gas)
arising from transitions between the symmetric and the antisymmetric
energy levels (i.e. the lowest two energy levels shown in Fig.\ 1) ---
the elementary excitations associated with transitions to the higher
subbands are at considerably higher energies and were not studied in
Ref.\ \onlinecite{decca}.
Thus, the SPE peak corresponding to the pole in $\Pi_0$ always occurs at
the energy $E=\bigtriangleup_{\text{SAS}}$ corresponding to the
(LDA-renormalized)
symmetric-antisymmetric gap between the lowest two levels.
Note that in our calculations of the spectral weight functions
($Im \; \Pi_0$, $Im \; \Pi$, $Im \; \tilde{\Pi}$ corresponding
respectively to SPE, SDE, and CDE) we use a small collisional broadening
$\Gamma \approx 0.1 meV$ taken from the experimental mobility values.

It is traditional to write the intersubband elementary excitation energies
as \cite{ando,mar-das,and-fow-ste,pinczuk,gam}

\begin{equation}
E_{\text{CD}}^2=E_{\text{SD}}^2+2 E_{\text{SP}}(n_{\text{S}}-n_{\text{AS}})
\alpha^\ast
\end{equation}

\begin{equation}
E_{\text{SD}}^2=E_{\text{SP}}^2-2 E_{\text{SP}}(n_{\text{S}}-n_{\text{AS}})
\beta^\ast
\end{equation}

\noindent where $n_{\text{S(AS)}}$ are the occupancies of the symmetric
(antisymmetric) subbands (i.e. $N_{\text{S}}=n_{\text{S}}+n_{\text{AS}}$),
$E_{\text{SP}} \equiv \bigtriangleup_{\text{SAS}}$, and $\alpha^\ast$,
$\beta^\ast$ are
parameters (which depend on $E_i$, $\phi_i(z)$, $V$, and $U_{\text{xc}}$)
which determine \cite{footnote} the depolarization shift and the vertex
correction, respectively.
Note \cite{footnote} that the definitions of the depolarization shift
($\alpha^\ast$)
and the vertex correction ($\beta^\ast$) shifts as given in Eqs.\ (1)
and (2) above explicitly incorporate the occupancy factor dependence
(i.e. the $n_{\text{S}}-n_{\text{AS}}$ factor) arising from the Pauli
principle, thus eliminating the trivial dependence of both the depolarization
shift and the vertex correction on the occupancy difference as both (S and AS)
subbands are occupied.
Any dependence of the depolarization shift and the vertex correction
parameters $\alpha^\ast$ and $\beta^\ast$ on the subband occupancy
factor $\eta$ (where $\eta<1$ means both subbands are occupied) necessarily
arises from nontrivial screening and exchange-correlation corrections and
{\em not} as a trivial manifestation of the Pauli principle.
Following the experimental procedure of Ref.\ \onlinecite{decca}, we calculate
the $\alpha^\ast$ and $\beta^\ast$ parameters by obtaining $E_{\text{SP}}$
(poles of $Im \; \Pi_0$, i.e. the LDA energy levels), $E_{\text{SD}}$
(poles of $Im \; \Pi$ including vertex corrections), and $E_{\text{CD}}$
(poles of $Im \; \tilde{\Pi}$) from our time-dependent LDA-linear
response calculations.
(The occupancies $n_{\text{S}}$ and $n_{\text{AS}}$ are known from our
LDA calculations.)

In Figs.\ 3 and 4 we show a direct comparison between our theoretical
calculations and experimental measurements for $\beta^\ast$ and $\alpha^\ast$
as functions of the filling parameter $\eta$ (Fig.\ 3) and the total
electron density $N_{\text{S}}$ (Fig.\ 4) for all seven samples employed
by Decca {\it et al.} \cite{decca}.
In general, $\beta^\ast$ goes down with decrease (increase) in $\eta$
($N_{\text{S}}$) both in experiment and theory, showing very good
agreement (better than 0.3 meV in absolute energies of
$\bigtriangleup_{\text{SAS}} - E_{\text{SD}}$)
except for one
qualitative difference, namely, in theory $\beta^\ast$ decreases
monotonically with decreasing (increasing) $\eta$ ($N_{\text{S}}$)
whereas in experiment $\beta^\ast$ seems to go abruptly to zero
around $\eta \approx 0.1$ (i.e. the vertex correction vanishes around
$\eta \approx 0.1$ making the SDE and SPE indistinguishable).
Note that even around $\eta \approx 0.1$, the actual energetic
difference between our theoretically predicted SDE peak and the
experimental ``SDE'' peak ($\equiv$ ``SPE'' with $\beta^\ast=0$)
is typically small ($\sim$ 0.2--0.3 meV).
For the depolarization shift ($\alpha^\ast$) experiment and theory,
in general, agree very well (except again for $\eta \stackrel{<}{\sim} 0.25$
there is some quantitative difference with the experimental CDE energies
being typically 0.5 meV below the theoretical calculations).
The important point to note about the depolarization shift is that
$\alpha^\ast$ is reasonably insensitive to variation in $\eta$
(or, $N_{\text{S}}$), changing little over an almost order of magnitude
change in electron density, both in experiment and theory.

For obvious reasons (namely, that it is not possible to vary $N_{\text{S}}$
or $\eta$ in a single DQW sample---each sample comes with its fixed
$N_{\text{S}}$ and $\eta$ values) the experimental results (and the
corresponding theoretical results) are for seven different DQW samples
corresponding to the seven different data points shown in Figs.\ 3 and 4.
Theoretically, of course, we can vary $\eta$ continuously in a fixed
sample by changing $N_{\text{S}}$ continuously in our calculations.
Results of such a calculation for a fixed sample are shown in Fig.\ 5
as continuous functions of $N_{\text{S}}$ (main figures) and $\eta$ (insets).
(For completeness, in Fig.\ 5 we show $\beta^\ast$ calculated using
both local-charge-density and local-spin-density exchange-correlation
potentials \cite{gam}. The difference between the two curves is less
than the experimental errors for the range of $\eta$ under consideration.)
These results clearly demonstrate (without the necessary numerical scatter
of Figs.\ 3 and 4 which arise from using different samples) that:
(i) The vertex correction $\beta^\ast$ decreases monotonically with
decreasing (increasing) $\eta$ ($N_{\text{S}}$), becoming very small as
$\eta$ approaches zero; (ii) The calculated $\beta^\ast$ does not agree
well with experiment for $\eta \stackrel{<}{\sim}0.25$, and most notably,
it does not go abruptly to zero for a finite value of $\eta$, as it does in
the experiment for $\eta \approx 0.1$; (iii) The depolarization
shift $\alpha^\ast$ is essentially insensitive to
$\eta$ (or $N_{\text{S}}$) remaining a
constant over a broad range of density and occupancy values.

In summary, we have theoretically calculated the lowest intersubband
SDE and CDE energies in a strongly-coupled two-component DQW
structure where both the symmetric and the antisymmetric subbands are
occupied, finding, in good agreement with a recent experiment \cite{decca},
that the vertex correction $\beta^\ast$ decreases monotonically with
decreasing subband occupancy parameter
$\eta \equiv \bigtriangleup_{\text{SAS}} /E_{\text{F}}$
or increasing the electron density $N_{\text{S}}$ whereas the depolarization
shift $\alpha^\ast$ is insensitive to changing $\eta$ and $N_{\text{S}}$.
In contrast to the experimental finding that $\beta^\ast \approx 0$
abruptly around $\eta \approx 0.1$, $\beta^\ast$ decreases monotonically
in our theory
with decreasing (increasing) $\eta$ ($N_{\text{S}}$) becoming very small
for small $\eta$.
Thus, while our time-dependent LDA theory correctly describes the broad
quantitative features of experiment quite well, the abrupt collapse of the
vertex correction must be arising from higher order vertex diagrams
(as discussed in Ref.\ \onlinecite{decca}) not included in the ladder vertex
corrections \cite{mar-das} of our time-dependent-LDA theory.
Our results for $\beta^\ast$ can be meaningfully understood as the screening
out of the excitonic correction in the high density limit which makes the
vertex correction vanish.
While the excitonic correction arising from exchange
interaction is screened, the direct Coulomb interaction leading to the
depolarization shift is obviously unscreened, and, therefore,
$\alpha^\ast$ remains unchanged as $N_{\text{S}}$ increases.

\section*{ACKNOWLEDGMENTS}

We thank Drs.\ R.\ Decca and A.\ Pinczuk for helpful discussions
and for communicating to us their unpublished data.
This work is supported by the US-ARO, US-ONR, and the NSF-MRG.

\begin{figure}
\caption{A typical double quantum well structure, given by the bare
confining potential $V_{\text{CONF}}$, used in the experimental
study of Ref.\ [1], and its self-consistent LDA subband energy levels $E_i$,
eigenfunctions $\phi_i$, electron density $n(z)$, Fermi energy $E_{\text{F}}$,
and effective, Hartree, and exchange-correlation potentials $V_{\text{EFF}}$,
$V_{\text{H}}$, and $V_{\text{XC}}$.
The areal density is $N_s=2.68 \times 10^{11} \; cm^{-2}$.}
\end{figure}

\begin{figure}
\caption{Calculated Raman spectra for the spin density excitation (SDE),
and the single particle excitation (SPE), given by $Im \; \Pi$ and
$Im \; \Pi_0$, respectively, for several values of the in-plane
wavevector transfer $q$.
The DQW sample has well width $d_{\text{w}}=139\text{\AA}$, barrier
width $d_{\text{b}}=28\text{\AA}$, density
$N_{\text{S}}=6.35\times10^{11}cm^{-2}$ and subband occupancy
parameter $\eta=0.12$.}
\end{figure}

\begin{figure}
\caption{Calculated and experimentally measured depolarization shift
and excitonic vertex correction parameters $\alpha^\ast$ and $\beta^\ast$
as functions of the filling parameter
$\eta \equiv \bigtriangleup_{\text{SAS}} / E_{\text{F}}$,
for the different DQW samples studied in the experiments of Ref.\ [1].}
\end{figure}

\begin{figure}
\caption{Calculated and experimentally measured depolarization shift
and excitonic vertex correction parameters $\alpha^\ast$ and $\beta^\ast$
as functions of the total electron density $N_{\text{S}}$,
for the different DQW samples studied in the experiments of Ref.\ [1].}
\end{figure}

\begin{figure}
\caption{Calculated depolarization shift and excitonic vertex correction
parameters $\alpha^\ast$ and $\beta^\ast$ as functions of areal density
$N_{\text{S}}$, for a DQW structure of well width 139$\text{\AA}$ and barrier
width 40$\text{\AA}$.
Insets: $\alpha^\ast$ and $\beta^\ast$ for the same DQW as functions of
the filling parameter
$\eta \equiv \bigtriangleup_{\text{SAS}} / E_{\text{F}}$.
The solid and dashed lines give $\beta^\ast$ calculated using
local-charge-density and local-spin-density exchange-correlation potentials,
respectively.}
\end{figure}


\begin{thebibliography}{99}
\bibitem{footnote} Following references
\onlinecite{ando,mar-das,and-fow-ste,pinczuk,gam}
one could define $\alpha=(n_{\text{S}}-n_{\text{AS}}) \alpha^\ast$
and $\beta=(n_{\text{S}}-n_{\text{AS}}) \beta^\ast$, which are more
appropriate in the extreme quantum limit ($\eta>1$) when only the lowest
subband is occupied, making $\alpha=N_{\text{S}} \alpha^\ast$
and $\beta=N_{\text{S}} \beta^\ast$.
When both subbands are occupied ($\eta<1$) it is appropriate to eliminate
the trivial Pauli principle-induced occupancies factor from the definitions
of $\alpha$ and $\beta$ as was done in Ref.\ \onlinecite{decca}.
\end{thebibliography}
\end{document}